\documentclass[pra,aps,showpas,floatfix,superscriptaddress,twocolumn]{revtex4}
\usepackage{graphicx}
\usepackage{bm} \usepackage{amssymb}
\usepackage{amsmath} \usepackage{amsfonts}

\newcommand{\ket}[1]{\left\vert#1\rangle\right.}
\newcommand{\bra}[1]{\left.\langle#1\right\vert}

\newcommand{\dtime}[1]{\frac{\partial#1}{\partial t}}
\newcommand{\I}{\dot{\imath}}

\begin{document}
\title{Quantum control of EIT dispersion via atomic tunneling in a double well Bose-Einstein condensate}
\author{James Owen Weatherall}
\affiliation{Department of Physics and Engineering Physics,
Stevens Institute of Technology, Castle Point on Hudson, Hoboken, NJ 07030, USA}
\affiliation{Department of Mathematical Sciences,
Stevens Institute of Technology, Castle Point on Hudson, Hoboken, NJ 07030, USA}
\affiliation{Department of Logic and Philosophy of Science,
University of California Irvine, 3151 Social Science Plaza A, Irvine, CA 92697, USA}
\author{Christopher P. Search}
\affiliation{Department of Physics and Engineering Physics,
Stevens Institute of Technology, Castle Point on Hudson, Hoboken, NJ 07030, USA}
\author{Markku J{\"a}{\"a}skel{\"a}inen}
\affiliation{Department of Physics and Engineering Physics,
Stevens Institute of Technology, Castle Point on Hudson, Hoboken, NJ 07030, USA}

\date{\today}

\begin{abstract}
Electromagnetically induced transparency (EIT) is an important tool for controlling light propagation and nonlinear wave mixing in atomic gases with potential applications ranging from quantum computing to table top tests of general relativity. Here we consider EIT in an atomic Bose-Einstein Condensate (BEC) trapped in a double well potential. A weak probe laser propagates through one of the wells and interacts with atoms in a three-level $\Lambda$ configuration. The well through which the probe propagates is dressed by a strong control laser with Rabi frequency $\Omega_{\mu}$, as in standard EIT systems.  Tunneling between the wells at the frequency $g$ provides a coherent coupling between identical electronic states in the two wells, which leads to the formation of inter-well dressed states. The macroscopic interwell coherence of the BEC wave function results in the formation of two ultra-narrow absorption resonances for the probe field that are inside of the ordinary EIT transparency window. We show that these new resonances can be interpreted in terms of the inter-well dressed states and the formation of a novel type of dark state involving the control laser and the inter-well tunneling.  To either side of these ultra-narrow resonances there is normal dispersion with very large slope controlled by $g$. For realistic values of $g$, the large slope of this dispersion yields group velocities for the probe field that are two orders of magnitude slower than standard EIT systems.  We discuss prospects for observing these ultra-narrow resonances and the corresponding regions of high dispersion experimentally.
\end{abstract}
\maketitle

\section{Introduction}

Electromagnetically induced transparency (EIT) \cite{harris-first-EIT} is a quantum interference effect that occurs in coherently prepared three-level $\Lambda$ atomic systems. The great utility of EIT comes from the fact that an ordinarily opaque medium can be made transparent to a probe laser while at the same time having large controllable dispersion and large third order nonlinear susceptibilities \cite{fleischhauer-review, scully-text}. EIT uses a strong control beam to dress an electronic excited state with a third auxilliary level. A weak probe field, which normally has only a single excitation path from the ground state to the excited state in the absence of the control beam, now has two excitation pathways to the excited state, corresponding to the two dressed states formed with the auxilliary state. The resulting destructive quantum interference between excitation pathways leads to vanishing absorption at the bare atomic resonance. Along with the vanishing of the probe absorption, the real part of the linear susceptibility, $\Re[\chi^{(1)}]$, exhibits normal dispersion with a very large slope leading to extremely slow group velocities for the probe field \cite{lukin+imamoglu, leonhardt-primer}. Slow light propagation through EIT systems has been observed experimentally in a variety of media, including hot atomic gases \cite{lukin+scully} and atomic Bose-Einstein condensates (BEC's) \cite{hau-first-observation}, and is now well understood.

Slow light propagation in EIT can be thought of in terms of the quasiparticles known as ``dark states polaritons'' \cite{polaritons, lukin-review}, which are a superposition of the probe pulse and the atomic polarization of the ground states. Beyond the novelty of simply controlling the speed of light, EIT has found numerous potential applications. Dark state polaritons provide a method for fully reversible storage of light pulses in an atomic medium by adiabatically switching on and off the control laser \cite{stopped-light}. Light storage has important applications for quantum information processing since quantum information can now be transmitted by flying qubits (photons) between stationary qubits (atomic ensembles) in a quantum network. Besides engineering the linear susceptibility, EIT results in constructive quantum interference for the nonlinear susceptibility, $\chi^{(3)}$, in the middle of the transparency window where the absorption vanishes and the dispersion is large. Such large nonlinearities in lossless media lead to an efficient scheme for four-wave mixing and frequency conversion in atomic vapors \cite{harris-first-EIT,zhang-1993,petch-1996}. Additional work has shown that these large nonlinearities can be used to achieve nonlinear mixing between pulses involving a few photons \cite{harris-lowlight}, which could be used to create an all-optical controlled-NOT gate \cite{schmidt}, the essential element of a quantum computer. Furthermore, one of the most striking applications of EIT has been the realization that the propagation of ultraslow light in moving atomic media is mathematically the same as light propagation in curved spacetime \cite{bathtub}. Leonhardt and Piwnicki \cite{leonhardt-piwnicki} showed that in this case a vortex, such as in an atomic Bose-Einstein condensate, will behave like a black hole for the light. This opens up the possibility of table-top tests of general relativity.

In the current paper, we describe a modification of the standard 3-level EIT configuration that utilizes coherent tunneling of a BEC in a double well potential and leads to qualitative changes in the linear susceptibility of the probe laser, which, as a result, provides additional control over the dispersion. More specifically, we consider the optical properties of an atomic BEC \cite{leggett-review} with three electronic states in a $\Lambda$-configuration that is trapped in a double well potential \cite{double-well-dynamics, josephson-junction,josephson-junction-APB,josephson-junction-nature}.  One well is prepared as a standard EIT system: the electronic excited state is coupled to one of the two stable ground states via a strong control laser, while a weak probe couples the other ground state to the same excited state. Both lasers are confined to a single well, leaving the second well unperturbed by them. However, the barrier between the wells represents a weak link through which atoms can tunnel. The global phase coherence of the condensate wave function can lead to phase coherent tunneling of the condensate wave function between the wells. This tunneling is the origin of Josephson oscillations of the population difference between the wells, which have recently been observed in a double well condensate \cite{josephson-junction,josephson-junction-APB,josephson-junction-nature}.

This double well BEC `Josephson junction' significantly modifies the probe EIT spectrum since the tunneling transforms the 3-level $\Lambda$ system of a lone well into a 6-level system spatially distributed between the two wells. The additional eigenstates of the 6-level system manifest themselves in the form of new absorption resonances in the probe susceptibility. In particular, we show here that the tunneling induces two qualitatively new ultranarrow absorption resonances situated in the middle the EIT transparency window with widths and positions determined by the tunneling frequencies between the wells. These new resonances are clearly visible when the tunneling frequencies are much less than the control laser Rabi frequency such that the transparency window is larger than the separation between these resonances. At the same time, the change in the probe index of refraction to either side of these new resonances is more dramatic still than found in a standard EIT system with a control laser of the same intensity, providing additional control over the group velocity and shape of a light pulse traveling through the medium. We predict that for realistic tunneling rates ($\sim 1$KHz), the dispersion to either side of the new resonance can yield group velocities up to two orders of magnitude slower than would be possible in EIT without the tunneling. Although one can in principle reduce the group velocity in EIT by simply turning down the intensity of the control laser, this approach narrows the transparency window and is fundamentally limited by ground state decoherence. By contrast, the ultranarrow tunneling resonances presented here are independent of the ground state decoherence and can be used to create ultra-slow group velocities independently of the control laser strength.


The remainder of this paper is organized as follows.  In section II, we will describe our model for the two-well $\Lambda$ BEC dressed by both the control beam and tunnel coupling, which has an analytic steady-state solution to its master equation.  In section III, we will derive the system's linear susceptibility, $\chi^{(1)}$, from which we can extract the absorption coefficient and dispersion.  In section \ref{experimental_prospects}, we will consider the prospects for experimental observation of the ultra-narrow features we describe, and their consequences for slow light.

\section{The Model}\label{the_model}

The present contribution concerns a gas of $N$ weakly interacting atoms of a Bose-Einstein condensate trapped in two neighboring wells of a double well potential.  There have been various experimental realizations of double well potentials for BEC's involving some combination of magnetic and/or optical dipole potentials. The first of these used a focused blue-detuned far-off resonant laser in the center a harmonic magnetic trap \cite{BEC-interference}. Later attempts created double well potentials via two parallel laser beams that generated adjacent optical dipole traps within the same condensate \cite{BEC-double-well-ketterle, BEC-double-well-ketterle-2}. In these cases, tunneling between wells was negligible. More recently, a double well potential with coherent quantum mechanical tunneling of the condensate wave function between wells was demonstrated \cite{josephson-junction,josephson-junction-APB,josephson-junction-nature}. These represented the first realizations of a single Josephson junction in an atomic BEC and serve as a guide for our EIT model, since coherent coupling of the wells is the essential new element. In these experiments \cite{josephson-junction-APB}, the double well was created by superimposing a one dimensional optical lattice on top of the harmonic optical dipole trap leading to a potential in the x-direction,
\begin{equation}
V_{\ell}(x)=\frac{1}{2}m\omega_{\ell} x^2 +V_{\ell}\cos^2\left(\frac{\pi x}{d_{\ell}}\right) \label{double well}
\end{equation}
where $d_{\ell}$ is the lattice constant and $\ell$ is the electronic state of the atoms.  This is because, in general, any magnetic or optical potentials used to trap the atoms will depend on their electronic state and therefore atoms in different states will experience slightly different trapping potentials.
We assume that in the z-direction, the harmonic trapping potential is much weaker than in the x or y directions, leading to an elongated cigar shaped potentials for the two wells with the long axis along the z-direction.

\begin{figure}
\includegraphics[width=.68\columnwidth,angle=270,viewport= 0 205 410 100]{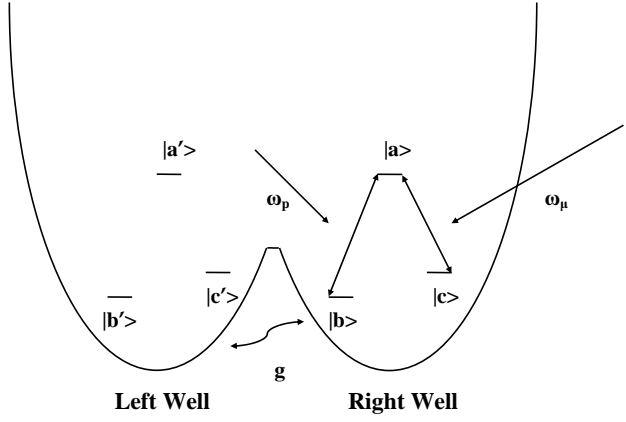}
\caption{\label{the-system}Schematic description of our system: the atoms in the right well are dressed by a strong control beam (indicated here by its frequency, $\omega_{\mu}$) near resonance with the $\ket{c}\rightarrow\ket{a}$ transition.  We are interested in the behavior of a weak probe beam (here, $\omega_p$) propagating in the right well, near resonance with the $\ket{b}\rightarrow\ket{a}$ transition. Atoms in electronic state $|\ell\rangle$ are coupled via tunneling through the inter-well barrier to the corresponding states $|\ell'\rangle$ in the left well.  The lasers propagate along the axis perpendicular to the page.}
\end{figure}

We consider three internal electronic states of the atoms in a $\Lambda$ configuration, denoted by eigenkets $|a\rangle$, $|b\rangle$, and $|c\rangle$ where $|a\rangle$ is an electronically excited state, while $|b\rangle$ and $|c\rangle$ are hyperfine ground states of the atoms. The direct transition between the two lower levels is assumed to be dipole forbidden while the transition between the highest level and each of the lower levels are allowed optical dipole transitions. Here we use a $'$ to denote the same internal states but in the left well so that for example, $|a\rangle$ is the electronic excited state in the right well while $|a'\rangle$ represents the same internal state of the atom but now that atom is located in the left well (see Fig. \ref{the-system}).

In analogy to the standard EIT configuration, we assume one of the two wells (in this case the right well) is dressed with a strong control beam with electric field amplitude $\mathcal{E}_{\mu}$ and frequency $\omega_\mu$ that is close to resonance with the energy difference between levels $\ket{a}$ and $\ket{c}$. Here we are concerned with the propagation through the right well of a weak probe field, $\mathcal{E}_{p}$, with frequency $\omega_p$ near resonance with the $\ket{b}\rightarrow\ket{a}$ transition.

The restriction that the lasers interact with only a single well should be achievable provided the spacing between the wells is sufficiently larger than the diffraction limit. The diffraction limit is essentially given by the wavelength of the probe and control lasers, which we denote simply as $\lambda$. Based on Eq. \ref{double well}, the well spacing must satisfy $d\gg \lambda$, which can be achieved with current technology. For example, in the experiment of Ref. \cite{josephson-junction} the spacing between the wells is $4.4\mu m$, which is significantly larger than a typical optical wavelength. Furthermore, the group in Ref. \cite{josephson-junction-nature} were able to optically resolve a single well to successfully image tunneling effects.  The probe and control lasers are assumed to propagate along the z-axis to maximize the optical thickness of the sample.

As we are working in the zero-temperature limit, we adopt the Hartree approximation and assume that all of the particles are co-existent in a single fully-condensed state. We model the wells as weakly coupled harmonic potentials \cite{double-well-dynamics} with ground state wave functions, $u^{(L/R)}_{\ell}({\bf r})$, localized in the left (L) or right (R) wells, which also depend on the electronic state since they represent the localized ground state near the minima of the state dependent potential, $V_{\ell}$. We assume that the overall condensate wave function, ${\bf \Psi}({\bf r},t)$ can be expressed in terms of these basis functions $u_{\ell}({\bf r})$,
\begin{equation}
\mathbf{\Psi}({\bf r},t)=\sqrt{N} \sum_{\ell=a,b,c}\psi_\ell(t) u^{(R)}_{\ell}({\bf r})\ket{\ell} +\sum_{\ell'=a',b',c'}\psi_{\ell'}(t) u^{(L)}_{\ell}({\bf r})\ket{\ell'}
\end{equation}
As a matter of notation, we introduce the vectors of probability amplitudes for the right and left wells, respectively, $\Psi = (\psi_a, \psi_b, \psi_c)^T$ and likewise $\Psi' = (\psi_{a'}, \psi_{b'}, \psi_{c'})^T$.

We work in a rotating frame defined by
\begin{align*}
\Psi \rightarrow \tilde{\Psi}=R \Psi;& &\Psi' \rightarrow \tilde{\Psi'}=R \Psi',
\end{align*}
where
\begin{equation}\label{rotating_frame}
R =\begin{bmatrix}
e^{\I(\omega_p+\omega_{\mu})t} &0 &0\\
0 &e^{\I \omega_{\mu}t} &0\\
0 &0 &e^{\I\omega_{p}t}
\end{bmatrix}.
\end{equation}
In this basis, the Gross-Pitaevskii equations for the six probability amplitudes are:
\begin{subequations}\label{gpes}
\begin{eqnarray}
\I \dtime{\tilde{\psi}_{a}} &= &\left(\omega_a-\omega_p-\omega_{\mu}+\tilde{\Psi}^{\dagger}\mathbf{U}_a\tilde{\Psi}\right)\tilde{\psi}_a - \frac{\Omega_{ab}}{2}e^{-\I\phi_{ab}}\tilde{\psi}_b \nonumber\\
&- &\frac{\Omega_{ac}}{2}e^{-\I\phi_{ac}}\tilde{\psi}_c - \frac{g_a}{2} \tilde{\psi}_{a'}\\
\I \dtime{\tilde{\psi}_{b}} &= &\left(\omega_b-\omega_{\mu}+\tilde{\Psi}^{\dagger}\mathbf{U}_b\tilde{\Psi}\right)\tilde{\psi}_b - \frac{\Omega_{ab}}{2}e^{\I\phi_{ab}}\tilde{\psi}_a\nonumber\\
&- &\frac{g_b}{2} \tilde{\psi}_{b'}\\
\I \dtime{\tilde{\psi}_{c}} &= &\left(\omega_c-\omega_{p}+\tilde{\Psi}^{\dagger}\mathbf{U}_c\tilde{\Psi}\right)\tilde{\psi}_c - \frac{\Omega_{ac}}{2}e^{\I\phi_{ac}}\tilde{\psi}_a \nonumber\\&-
&\frac{g_c}{2} \tilde{\psi}_{c'}\\
\I \dtime{\tilde{\psi}_{a'}} &= &\left(\omega_a-\omega_{\mu}-\omega_{p}+\tilde{\Psi'}^{\dagger}\mathbf{U}_a\tilde{\Psi'}\right)\tilde{\psi}_{a'} - \frac{g_a}{2} \tilde{\psi}_{a}\\
\I \dtime{\tilde{\psi}_{b'}} &= &\left(\omega_b-\omega_{\mu}+\tilde{\Psi'}^{\dagger}\mathbf{U}_b\tilde{\Psi'}\right)\tilde{\psi}_{b'} - \frac{g_b}{2} \tilde{\psi}_{b}\\
\I \dtime{\tilde{\psi}_{c'}} &= &\left(\omega_c-\omega_p+\tilde{\Psi'}^{\dagger}\mathbf{U}_c\tilde{\Psi'}\right)\tilde{\psi}_{c'} - \frac{g_c}{2} \tilde{\psi}_{c}.
\end{eqnarray}
\end{subequations}
We have assumed that these amplitudes are normalized to 1:
\[
\sum_{\ell=a,b,c}|\tilde{\psi}_\ell(t)|^2+\sum_{\ell=a',b',c'}|\tilde{\psi}_\ell(t)|^2=1
\]
Here we have incorporated the ground state energies of atoms in the wells, $\int d^3r[u^{(k)}_{\ell}({\bf r})]^*[-\hbar^2\nabla^2/2m+V_{\ell}({\bf r})]u^{(k)}_{\ell}({\bf r})$, into the definition of the atomic energy levels, $\omega_{\ell}$. The couplings between levels are moderated by their complex Rabi frequencies defined as $\hbar\Omega_{ac}e^{-\I \phi_{ac}} =\mathcal{E}_{\mu}  D_{ac}$ for the control field and $\hbar\Omega_{ab}e^{-\I \phi_{ab}} =\mathcal{E}_{p}  D_{ab}$ for the probe field. Here $\Omega_{ij}$ is taken to be real and $D_{ij}=e\bra{i}{\bf x}\cdot\epsilon\ket{j}$ is the dipole moment matrix element in the direction of the laser polarization, $\epsilon$.

In principle, each of the atomic levels is subject to a different coupling constant for the tunneling between wells.  We denote these couplings by $\hbar g_{\ell}/2=-\int d^3r [u^{(L)}_{\ell}({\bf r})]^*[-\hbar^2\nabla^2/2m+V_{\ell}({\bf r})] u^{(R)}_{\ell}({\bf r})$. For the sake of completeness, we note that Ref. \cite{bergmann} shows that $\hbar g_{\ell}$ is equal to the Josephson coupling energy, $E_J$, that appears in the Hamiltonian for the bosonic Josephson junction \cite{gati-noise-thermometry}. The two-body interactions are denoted by a rank 3 tensor, $\mathbf{U}_{ijk}$, defined by:
\begin{equation}
\mathbf{U}_{i}=\begin{bmatrix}
U_{ia} &0 &0\\
0 &U_{ib} &0\\
0 &0 &U_{ic}
\end{bmatrix},
\end{equation}
where the index $i$ runs likewise over $a$, $b$, and $c$. The elements $U_{ij}=(4\pi\hbar a_{ij}N/m)\int d^3r |u^{(k)}_i({\bf r})|^2|u^{(k)}_j({\bf r})|^2$ represent the interaction strength (in units $s^{-1}$) between states $\ket{i}$ and $\ket{j}$ in the same well in terms of the s-wave scattering length between the two states, $a_{ij}$.

The absorptive and dispersive properties of the medium with respect to the probe are given by the linear susceptibility, $\chi^{(1)}$, which we can derive from the coherence between states $\ket{a}$ and $\ket{b}$. To proceed further we must introduce the density matrix $\rho$, defined as the outer product of the probability amplitudes,
\begin{equation}
\rho=(\Psi, \Psi')(\Psi,\Psi')^{\dagger}.
\end{equation}
For the sake of clarity we note that $(\Psi, \Psi')=(\psi_a, \psi_b, \psi_c,\psi_{a'}, \psi_{b'}, \psi_{c'})$.
By the product rule, we arrive at the rate of change of the density matrix:
\begin{equation}
\dtime{\rho}=\dtime{(\tilde{\Psi},\tilde{\Psi}')}(\tilde{\Psi},\tilde{\Psi}')^{\dagger}+(\tilde{\Psi},\tilde{\Psi}')\dtime{(\tilde{\Psi},\tilde{\Psi}')^{\dagger}}-\frac{1}{2}\{\Gamma,\rho\}.
\end{equation}
To incorporate decay, we have introduced the decay matrix $\Gamma$ whereby each element of the density matrix decays at the rate $\dot{\rho}_{ij}\propto -\gamma_{ij}\rho_{ij}$ where $\gamma_{ij}=(\gamma_i+\gamma_j)/2+\gamma^{(dp)}_{ij}$. Here, $\gamma_i$ for $i=a,b,c,a',b',c'$ are the decay rates for the populations and $\gamma^{(dp)}_{ij}$ is decoherence due to pure dephasing for $i\neq j$.  Extending our notation, we write the density matrix in the rotating frame defined in equation Eq. \ref{rotating_frame} as $\tilde{\rho}$.  Extracting the equation of motion for $\tilde{\rho}_{ab}$, we find:
\begin{eqnarray}\label{first-rhoab}
\I \dtime{\tilde{\rho}_{ab}}
&= & (\Delta_{p} -\I\gamma_{ab}+
\tilde{\Psi}^{\dagger}(\mathbf{U}_a-\mathbf{U}_b)\tilde{\Psi})\tilde{\rho}_{ab}\nonumber\\
&+ &\frac{\Omega_{ab}}{2}e^{-\I\phi_{ab}}(\tilde{\rho}_{aa}-\tilde{\rho}_{bb})- \frac{\Omega_{ac}}{2}e^{-\I\phi_{ac}}\tilde{\rho}_{cb}\nonumber\\ &+ &\frac{ g_b}{2}\tilde{\rho}_{ab'}
- \frac{g_a}{2}\tilde{\rho}_{a'b},
\end{eqnarray}
where we have defined the probe's detuning from the $\ket{a}\rightarrow\ket{b}$ transition, $\Delta_{p}=\omega_a-\omega_b-\omega_p$.

Before proceeding further, we note that equation (\ref{first-rhoab}) depends on five additional, mutually dependent linked differential equations, each of which is likewise coupled to other terms in the density matrix.  But we are only interested in the linear susceptibility for the probe field, so we may solve the resulting coupled system of equations to first order in the strength of the probe field $\mathcal{E}_p$, which we have assumed to be weak.  We assume that initially all of the atoms are in the two states $|b\rangle$ and $|b'\rangle$. As a result, up to order $\mathcal{E}_p^2$, we have $\tilde{\rho}_{aa}=\tilde{\rho}_{a'a'}=\tilde{\rho}_{aa'}=\tilde{\rho}_{a'a}=0$. Additionally, $|c\rangle$ only develops population at order $\mathcal{E}_p^2\mathcal{E}_{\mu}^2$ in perturbation theory and therefore to first order in the probe laser, $\tilde{\rho}_{cc}=\tilde{\rho}_{c'c'}=\tilde{\rho}_{ac}=\tilde{\rho}_{a'c}=\tilde{\rho}_{ac'}=\tilde{\rho}_{a'c'} =\tilde{\rho}_{cc'}=\tilde{\rho}_{c'c}=0$.  The only terms in the density matrix that are nonzero to zeroth order in the probe are $\tilde{\rho}_{bb}$, $\tilde{\rho}_{b'b'}$, $\tilde{\rho}_{b'b}$ and $\tilde{\rho}_{bb'}$ while to first order in the probe $\tilde{\rho}_{ab}$, $\tilde{\rho}_{a'b}$, $\tilde{\rho}_{a'b'}$, $\tilde{\rho}_{ab'}$, $\tilde{\rho}_{cb}$, $\tilde{\rho}_{c'b}$, $\tilde{\rho}_{cb'}$, and $\tilde{\rho}_{c'b'}$ are nonzero.

The control laser is assumed to be of arbitrary strength so that we must solve the equations to all orders in $\Omega_{ac}$. In addition to this, we solve to all orders in the tunneling rates $g_{\ell}$. This is because the critical element of EIT is the presence of coherence between the two ground states, $\rho_{cb}$. In the case that the tunnel coupled states $\ket{b}$ and $\ket{b'}$ as well as $|c\rangle$ and $|c'\rangle$ are nearly degenerate, they will form superposition states between the two wells that will in turn affect $\rho_{cb}$. It is important to point out that our choice to include $g_b$ and $g_c$ to all orders is not at odds with our choice to only keep terms to linear order in the probe despite the fact that $g_b$ and $g_c$ are themselves small. The assumption of a weak probe means that $\rho_{aa}\ll \rho_{bb}$ at all times. To second order in perturbation theory one can easily show that $\rho_{aa}\sim (\Omega_{ab}/\gamma_a)^2$. Tunneling results in finite populations for both wells and in the case of degenerate states (including mean field interactions), there is equal population in both wells $\rho_{bb}=\rho_{b'b'}=1/2$. Consequently, as long as there is finite population in $|b\rangle$, the weak probe condition remains $\Omega_{ab}\ll \gamma_a$ and is only weakly effected by $g_i$. Appendix B gives the full solution for $\tilde{\rho}_{aa}$ to second order in $\Omega_{ab}$.

In order to keep the inter-well couplings to all orders, we move to a partially dressed state basis, in which the $\{\ket{b},\ket{b'}\}$ and $\{\ket{c},\ket{c'}\}$ subspaces of our effective Hamiltonian are diagonalized.  To simplify matters, we take $\gamma^{(dp)}=\gamma_b=\gamma_{b'}=\gamma_c=\gamma_{c'}=0$, which is a reasonable approximation because the decay rates for atoms in a BEC are given by the lifetime of the condensate, which is much longer than all other time scales in this problem. We keep the decay from the excited electronic state, $\gamma_a=\gamma_{a'}\neq 0$, which is due to spontaneous emission.

The effective Hamiltonian for the $\{\ket{b},\ket{b'}\}$ subspace can be written as a sum of its diagonal and traceless parts:
\begin{eqnarray}
\mathcal{H}_{bb'}&=&\hbar\left(\omega_{b}-\omega_{\mu}+\frac{1}{2}(\tilde{\Psi}^{\dagger}\mathbf{U}_b\tilde{\Psi}+\tilde{\Psi'}^{\dagger}\mathbf{U}_b\tilde{\Psi'})\right) \mathbf{I}\nonumber\\
&+&\frac{\hbar}{2}\begin{pmatrix}
\delta_{bb'} &-g_b\\
-g_b &-\delta_{bb'}
\end{pmatrix},
\end{eqnarray}
where $\delta_{bb'}=\tilde{\Psi}^{\dagger}\mathbf{U}_b\tilde{\Psi}-\tilde{\Psi'}^{\dagger}\mathbf{U}_b\tilde{\Psi'}$ is the energy difference between the corresponding states in each well.  In the case that $\rho_{bb}=\rho_{b'b'}$, $\delta_{bb'}=0$. If the wells are initially prepared with equal population in both of them, then $\delta_{bb'}=0$ initially and will remain zero since the eigenstates of $\mathcal{H}_{bb'}$ have equal probability to be $|b\rangle$ and $|b'\rangle$ in this case.

The matrix diagonalizing $\mathcal{H}_{bb'}$ will be a member of SO(2), which can be written in terms of a rotation angle in the $\{\ket{b},\ket{b'}\}$ subspace:
\begin{equation}
D_b=\begin{pmatrix}
\cos \theta_b &\sin \theta_b\\
-\sin \theta_b &\cos\theta_b
\end{pmatrix},
\end{equation}
where
\begin{subequations}
\begin{eqnarray}
 &\cos\theta_b = \left(\frac{1-\delta_{bb'}/g^{\text{eff}}_{b}}{2}\right)^{1/2}\\
&\sin\theta_b = \left(\frac{1+\delta_{bb'}/g^{\text{eff}}_{b}}{2}\right)^{1/2}\\
 &g_b^{\text{eff}}=\sqrt{\delta_{bb'}^2+g_b^2}.
\end{eqnarray}
\end{subequations}
Under this transformation, we find dressed states $\ket{B}$ and $\ket{B'}$ whose probability amplitudes can be written in terms of the bare states and the angle of rotation $\theta_b$:
\begin{subequations}
\begin{eqnarray}
\tilde{\psi}_B&=&\cos\theta_b\tilde{\psi}_b+\sin\theta_b\tilde{\psi}_{b'}\\
\tilde{\psi}_{B'}&=&-\sin\theta_b\tilde{\psi}_b+\cos\theta_b\tilde{\psi}_{b'}
\end{eqnarray}
\end{subequations}
Identical reasoning applies for the $\{\ket{c},\ket{c'}\}$ subspace leading to the dressed states $\{\ket{C},\ket{C'}\}$. Note that these dressed states are in a coherent super-position of spatially delocalized states.

Combining these transformations, we arrive at the full transformation from the bare basis $\{\ket{a}, \ket{b}, \ket{c},\ket{a'},\ket{b'}, \ket{c'}\}$
to the dressed basis $\{\ket{a}, \ket{B}, \ket{C},\ket{a'},\ket{B'}, \ket{C'}\}$,
\begin{equation}
D=\begin{pmatrix}
1 &0 &0 &0 &0 &0\\
0 &\cos\theta_b &0 &0 &\sin\theta_b &0\\
0 &0 &\cos\theta_c &0 &0 &\sin\theta_c\\
0 &0 &0 &1 &0 &0\\
0 &-\sin\theta_b &0 &0 &\cos\theta_b &0\\
0 &0 &-\sin\theta_c &0 &0 &\cos\theta_c
\end{pmatrix}.
\end{equation}
The Gross-Pitaevskii equations rewritten in the dressed basis are given in appendix A.

The transformation for the density matrix from the original basis, $\tilde{\rho}$, to the dressed basis is
$\tilde{\rho}_d=D\tilde{\rho} D^{\dagger}$. In terms of the dressed states, the coherence $\tilde{\rho}_{ab}$ is given by
\begin{equation}\label{rhoab}
\tilde{\rho}_{ab}=\cos\theta_b\tilde{\rho}_{aB}-\sin\theta_b\tilde{\rho}_{aB'}.
\end{equation}
Our assumption that all atoms are initially in some combination of $\{|b\rangle, |b'\rangle\}$, implies that in the dressed basis the terms $\rho_{BB}$, $\rho_{B'B'}$, and $\rho_{B'B}=(\rho_{BB'})^*$ are in general nonzero to zeroth order in the probe.
Beginning with these and keeping only terms to first order in $\Omega_{ab}$, we arrive at two decoupled systems of four equations each:
\begin{subequations}\label{system1}
\begin{eqnarray}
\I\dtime{\tilde{\rho}_{aB}}&=&\left(\Delta_p+\frac{g_b^{\text{eff}}}{2}-\frac{1}{2}U_{bb}+U_{ab}-\I\gamma_{ab}\right)\tilde{\rho}_{aB} \nonumber\\
&-& \frac{\Omega_{ac}}{2}e^{-\I\phi_{ac}}\left(\cos{\theta_c}\tilde{\rho}_{CB}-\sin{\theta_c}\tilde{\rho}_{C'B}\right)-\frac{g_a}{2}\tilde{\rho}_{a'B}\nonumber\\
&-&\frac{\Omega_{ab}}{2}e^{-\I\phi_{ab}}\left(\cos\theta_b\tilde{\rho}_{BB}-\sin\theta_b\tilde{\rho}_{B'B}\right)\\
\I\dtime{\tilde{\rho}_{CB}}&=&\left(\Delta_p-\Delta_{\mu}+\frac{g_b^{\text{eff}}}{2}-\frac{g_c^{\text{eff}}}{2}+\frac{1}{2}U_{cb}-\frac{1}{2}U_{bb}\right)\tilde{\rho}_{CB}\nonumber\\
&-&\frac{\Omega_{ac}}{2}e^{\I\phi_{ac}}\cos{\theta_c}\tilde{\rho}_{aB}\\
\I\dtime{\tilde{\rho}_{C'B}}&=&\left(\Delta_p-\Delta_{\mu}+\frac{g_b^{\text{eff}}}{2}+\frac{g_c^{\text{eff}}}{2}+\frac{1}{2}U_{cb}-\frac{1}{2}U_{bb}\right)\tilde{\rho}_{C'B}\nonumber\\
&+&\frac{\Omega_{ac}}{2}e^{\I\phi_{ac}}\sin{\theta_c}\tilde{\rho}_{aB}\\
\I\dtime{\tilde{\rho}_{a'B}}&=&\left(\Delta_p+\frac{g_b^{\text{eff}}}{2}-\frac{1}{2}U_{bb}+U_{ab}-\I\gamma_{ab}\right)\tilde{\rho}_{a'B}\nonumber\\
&-&\frac{g_a}{2}\tilde{\rho}_{aB};
\end{eqnarray}
\end{subequations}
and likewise
\begin{subequations}\label{system2}
\begin{eqnarray}
\I\dtime{\tilde{\rho}_{aB'}}&=&\left(\Delta_p-\frac{g_b^{\text{eff}}}{2}-\frac{1}{2}U_{bb}+U_{ab}-\I\gamma_{ab}\right)\tilde{\rho}_{aB'} \nonumber\\
&-& \frac{\Omega_{ac}}{2}e^{-\I\phi_{ac}}\left(\cos{\theta_c}\tilde{\rho}_{CB'}-\sin{\theta_c}\tilde{\rho}_{C'B'}\right)-\frac{g_a}{2}\tilde{\rho}_{a'B'}\nonumber\\
&+&\frac{\Omega_{ab}}{2}e^{-\I\phi_{ab}}\left( \sin\theta_b\tilde{\rho}_{B'B'}-\cos\theta_b\tilde{\rho}_{BB'}\right)\\
\I\dtime{\tilde{\rho}_{CB'}}&=&\left(\Delta_p-\Delta_{\mu}-\frac{g_b^{\text{eff}}}{2}-\frac{g_c^{\text{eff}}}{2}+\frac{1}{2}U_{cb}-\frac{1}{2}U_{bb}\right)\tilde{\rho}_{CB'}\nonumber\\
&-&\frac{\Omega_{ac}}{2}e^{\I\phi_{ac}}\cos{\theta_c}\tilde{\rho}_{aB'}\\
\I\dtime{\tilde{\rho}_{C'B'}}&=&\left(\Delta_p-\Delta_{\mu}-\frac{g_b^{\text{eff}}}{2}+\frac{g_c^{\text{eff}}}{2}+\frac{1}{2}U_{cb}-\frac{1}{2}U_{bb}\right)\tilde{\rho}_{C'B'}\nonumber\\
&+&\frac{\Omega_{ac}}{2}e^{\I\phi_{ac}}\sin{\theta_c}\tilde{\rho}_{aB'}\\
\I\dtime{\tilde{\rho}_{a'B'}}&=&\left(\Delta_p-\frac{g_b^{\text{eff}}}{2}-\frac{1}{2}U_{bb}+U_{ab}-\I\gamma_{ab}\right)\tilde{\rho}_{a'B'}\nonumber\\
&-&\frac{g_a}{2}\tilde{\rho}_{aB'}.
\end{eqnarray}
\end{subequations}
where the control laser detuning is $\Delta_{\mu}=\omega_a-\omega_c-\omega_{\mu}$. At this point we assume that the zeroth order populations in the dressed states are nonzero and controlled by a tunable parameter, $\varphi$, such that $\tilde{\rho}^{(0)}_{BB}=\cos^2(\theta_b-\varphi)$ and $\tilde{\rho}^{(0)}_{B'B'}=\sin^2(\theta_b-\varphi)$.  At the same time, we assume that coherences between the dressed states are initially zero, $\tilde{\rho}^{(0)}_{B'B}=\tilde{\rho}^{(0)}_{BB'}=0$. This is a reasonable assumption since if the atoms are specifically prepared at some time in the past in the dressed states or simply allowed to equilibrate to the eigenstates of the double well, then any coherences would be destroyed before the experiment by even a small amount of decoherence. The effect of initial coherences between dressed states on the transient probe absorption spectrum in a three level system has been considered before  \cite{berman} and shown to give rise to temporal oscillations in the absorption coefficient similar to optical nutation.

The structures of the solutions to these are identical.  In both cases, we write the systems as linear equations of the form
$\partial X/\partial t=-\mathbf{M}\cdot X(t)+A$,
and note that such equations have steady-state solutions of $\lim_{t\rightarrow\infty}X(t)=\mathbf{M}^{-1}\cdot A$.
The necessary terms $\tilde{\rho}_{aB}$ and $\tilde{\rho}_{aB'}$ are then the corresponding elements of the resulting vectors.  The general form of the analytic solution is quite complicated and is given in appendix B.

\section{Optical Properties of the Right Well in the Degenerate Energy Case}

The polarization for the probe is related to $\rho_{ab}$ by
\begin{equation}\label{genpol}
\mathcal{P}=2 N [u_a(\mathbf{r})]^*u_b(\mathbf{r}) D_{ab}\tilde{\rho}_{ab}.
\end{equation}
Likewise, the complex linear susceptibility is given by $\chi^{(1)}=\mathcal{P}/(\epsilon_0\mathcal{E}_p)$, which determines both the absorption coefficient, $\alpha(\omega_p)=k_p\Im[\chi^{(1)}]$ and the index of refraction, $n(\omega_p)\approx(1+\Re[\chi^{(1)}])^{1/2}$. The spatial term $[u_a(\mathbf{r})]^*u_b(\mathbf{r})$ reflects the density profile of the atoms and only determines the optical thickness of the condensate.

The simplest case to consider is when the dressed state mixing angles are $\theta_b=\theta_c=\pi/4$, corresponding to symmetric and antisymmetric superpositions between the two wells. This occurs when $\delta_{bb'}=\delta_{cc'}=0$ or equivalently when $\rho_{bb}=\rho_{b'b'}$.
In this case the solution simplifies considerably and the polarization is given by
\begin{equation}
\mathcal{P}(\Delta_p)=\frac{N [u_a(\mathbf{r})]^*u_b(\mathbf{r})D_{ab}^2 \mathcal{E}_p}{\hbar}\left(Z_++Z_-\right)\label{polarization-resonance}
\end{equation}
where
\begin{widetext}
\[
Z_{\pm}= \frac{\left((2 \Delta_{\mu}-2 \Delta_p\pm g_b)^2-g_c^2\right) (2 \Delta_p\mp g_b-2i\gamma_{ab})(1\mp\sin2\varphi)}{\left((2 \Delta_{\mu}-2 \Delta_p\pm g_b)^2-g_c^2\right) \left((2\Delta_p\mp g_b-2 i\gamma_{ab})^2-g_a^2\right)+(2 \Delta_{\mu}-2 \Delta_p\pm g_b) (2 \Delta_p\mp g_b- 2i\gamma_{ab}) \Omega_{ac}^2}
\]
\end{widetext}
Note that here we have redefined the probe and control laser detunings to include the mean field energy shifts,
$\Delta_p+U_{ab}\rightarrow \Delta_p$ and $\Delta_\mu +U_{ab} \rightarrow  \Delta_{\mu}$.  We emphasize the fact that, in the limit that $g_a,g_b,g_c,\varphi\rightarrow0$, we recover the standard EIT coherence, which for $\Delta_{\mu}=0$ has the form
\[
\tilde{\rho}_{ab}\rightarrow\frac{\Delta_p\Omega_{ab}e^{-\I\phi_{ab}}}{2\left(\Delta_p(\Delta_p-\I\gamma_{ab})-(\Omega_{ac}/2)^2\right)}.
\]

\begin{figure}
\includegraphics[width=1.0\columnwidth]{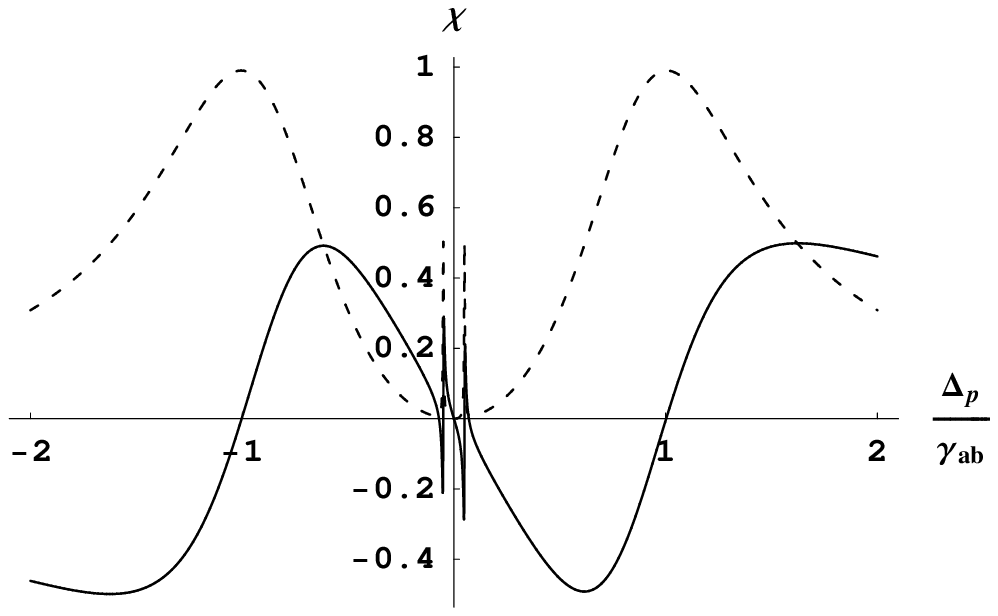}
\caption{\label{fig-spectrum}The full EIT spectrum of the system near the $\ket{a}-\ket{b}$ resonance.  The $\Im[\chi^{(1)}]$ is plotted as a dotted line, and $\Re[\chi^{(1)}]$ as a solid line.  Note the two additional features, symmetrically located around zero detuning, at $\pm g_b/2$, with equal amplitude for $\varphi=0$.  In this plot we have taken $g_b=g_c=\gamma_{ab}/10$ ($\approx 500KHz$) to emphasize the modifications to the standard EIT spectrum.  Given more physically realistic parameters, the features would be considerably narrower and closer together.  Cf. Figs. \ref{fig-resonance}, \ref{fig-varygc}, and \ref{fig-varygb}. Note that here and in subsequent figures $\Omega_{ac}=\gamma_a=2\gamma_{ab}$.}
\includegraphics[width=1.0\columnwidth]{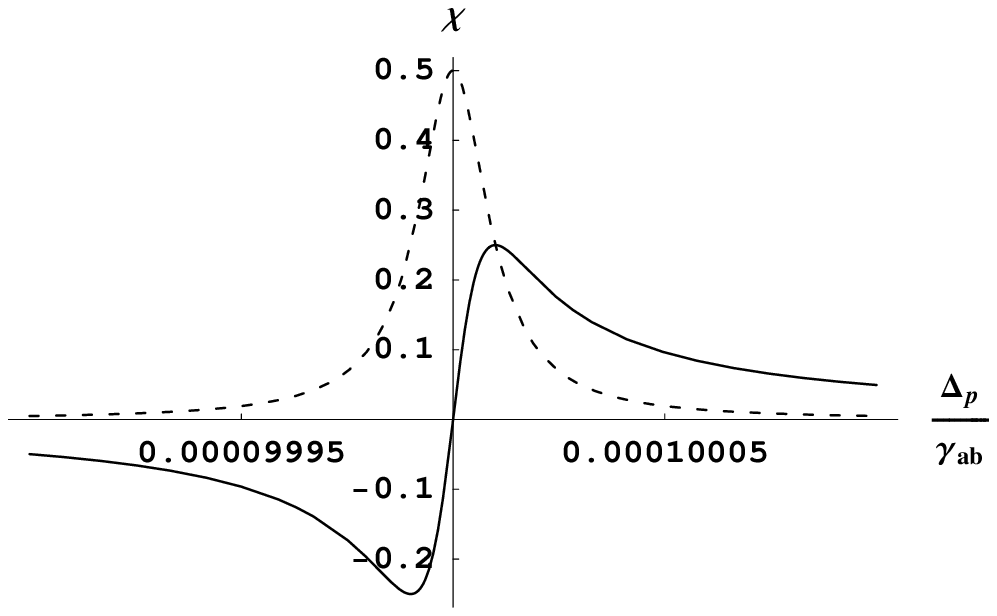}
\caption{\label{fig-resonance}Close-up of the positive detuning narrow absorption peak corresponding to the presence of the second well.  As in the previous figure, $\Im[\chi^{(1)}]$ is plotted as a dotted line, and $\Re[\chi^{(1)}]$ as a solid line. Note that to either side of the absorption peak, the real part of the linear susceptibility is rapidly changing, whereas the absorption goes to zero still more quickly.  In this plot, $g_b=g_c=(2\gamma_{ab})\times10^{-4}$, which we estimate as a reasonable upper bound for the coupling between wells (see text). Here $\varphi=0$.}
\end{figure}

\begin{figure}
\includegraphics[width=1.0\columnwidth]{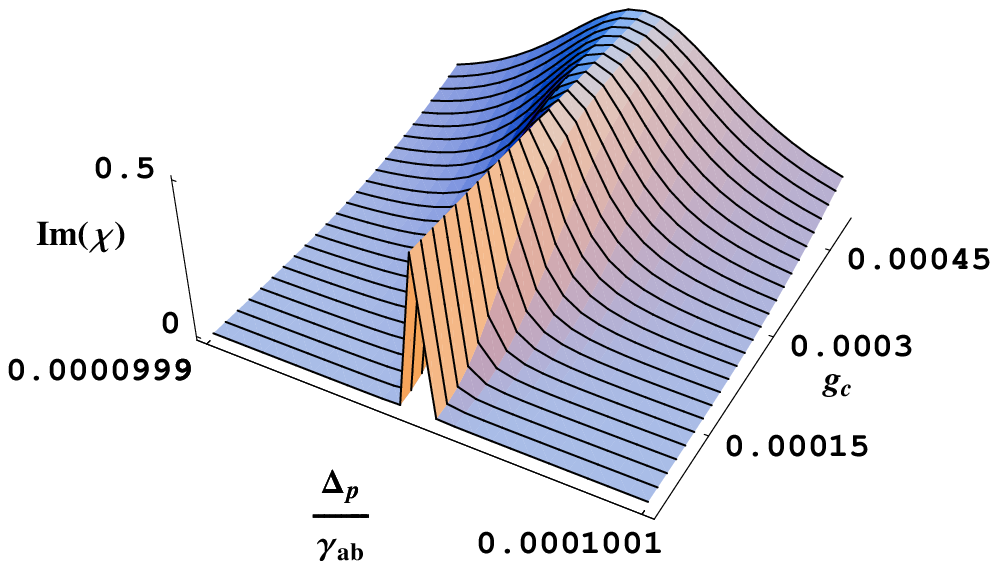}
\caption{\label{fig-varygc}$\Im[\chi^{(1)}]$, which is proportional to the probe absorption coefficient, plotted near the positive detuning resonance. Here $\varphi=0$, $g_b=(2\gamma_{ab})\times10^{-4}$ is fixed, and $g_c$ is varied, to demonstrate how $g_c$ moderates the width of the tunneling induced resonances.  See Fig. \ref{fig-resonance} for a cross-section of this plot at $g_c=1KHz$.}
\includegraphics[width=1.0\columnwidth]{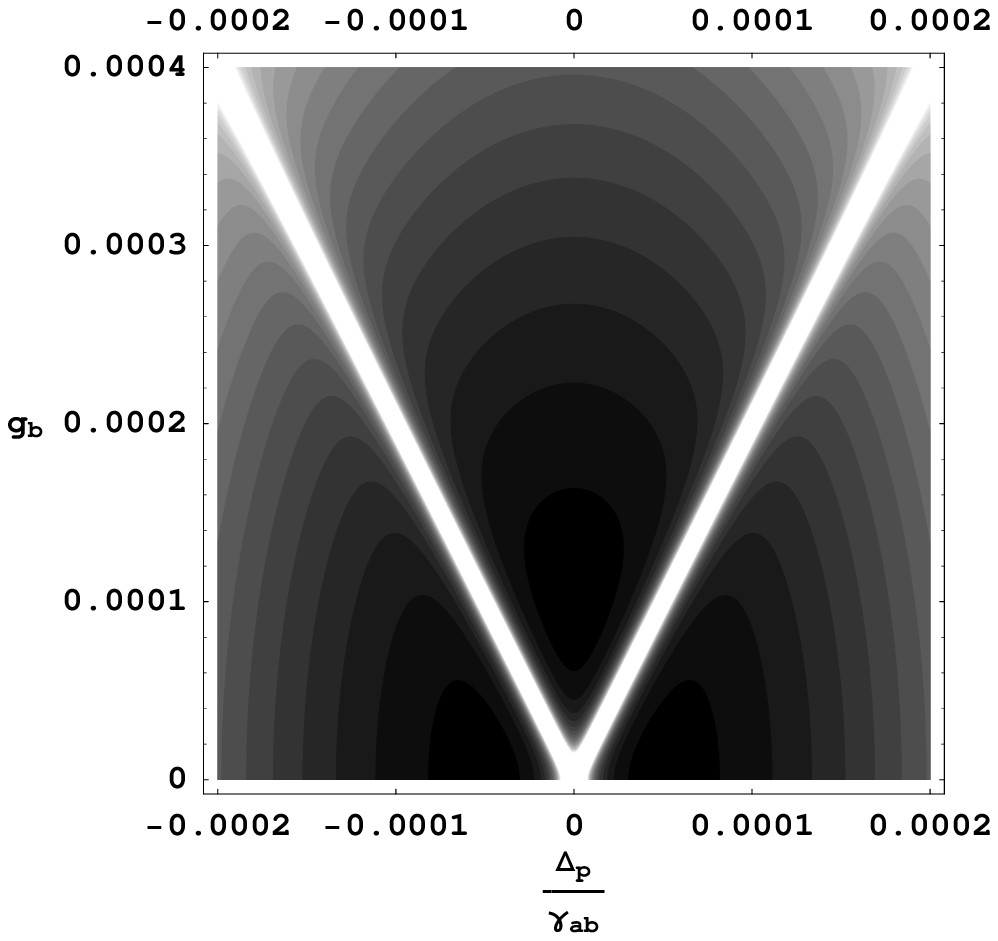}
\caption{\label{fig-varygb}$\Im[\chi^{(1)}]$ but now $g_b$ is varied, keeping $g_c=(2\gamma_{ab})\times10^{-4}$ fixed along with $\varphi=0$.  $g_b$ moderates the distance between the peaks.  As $g_b$ goes to zero, the peaks merge to create a single narrow peak at the origin.}
\end{figure}

Further analysis requires us to estimate the values for the important variables in the problem.  Atomic spontaneous emission rates for atoms commonly used in BEC experiments (Na, Rb, Li) are typically on the order of 10Mhz.  Typical tunneling times were on the order of $10ms$ in Ref. \cite{josephson-junction}
while more recent experiments achieved tunnel couplings between the wells as high 7900Hz \cite{gati-noise-thermometry}.  We therefore assume that 1KHz is a reasonable estimate for the coupling strength between two wells of a BEC. Therefore, unless stated otherwise, we will use the values $g_i=10^3 s^{-1}$ and $\gamma_a=10^7 s^{-1}$,  which yields $g_i=10^{-4} \gamma_a$. Examples of the real and imaginary parts of the susceptibility are shown in Figs. 2 and 3 for $\Delta_{\mu}=0$ while Figs. \ref{fig-varygc} and \ref{fig-varygb} display the spectrum's dependence on the tunneling parameters. (Note that in these and all subsequent figures the susceptibility is plotted in units of $N [u_a(\mathbf{r})]^*u_b(\mathbf{r})D_{ab}^2 /2\epsilon_0\hbar\gamma_{ab}$.) We see that the presence of the second well manifests itself as two ultranarrow resonances located inside of the EIT transparency window. When $\delta_{bb'}=0$, the new resonances are symmetrically located about $\Delta_p=0$ at the locations $\pm g_b/2$. In general the location of these resonances is $\Delta_p=\pm g_b^{eff}/2$ and for $\Omega_{ac},\gamma_{ab}\gg g_b,g_c$ their shape is approximately Lorentzian with a full width at half max of
\begin{equation}
\Gamma_n=2\left(\frac{g_c}{\Omega_{ac}}\right)^2\gamma_{ab} \label{linewidth}
\end{equation}
Note that we have also obtained this same result even when we have included decoherence due to pure dephasing between the ground states $|b\rangle \leftrightarrow |b'\rangle$, $|c\rangle \leftrightarrow |c'\rangle$, and $|b\rangle \leftrightarrow |c\rangle$ \cite{search-weatherall}. This implies that these resonances will be clearly separated even in the presence of finite ground state decoherence, and, in particular, $\gamma_{bc}$. Similar results have been obtained by Lukin {\em et al.} \cite{lukin-dark-resonances} and Mahmoudi {\em et al.} \cite{mahmoudi-dark-resonances} who studied EIT in 4-level system where an additional ground state was coupled via an RF or optical transition to the same ground state that is coupled to the control laser. They found an ultra-narrow resonance in the EIT spectrum with a line width of the same form as Eq. \ref{linewidth}.

\begin{figure}
\includegraphics[width=1.0\columnwidth]{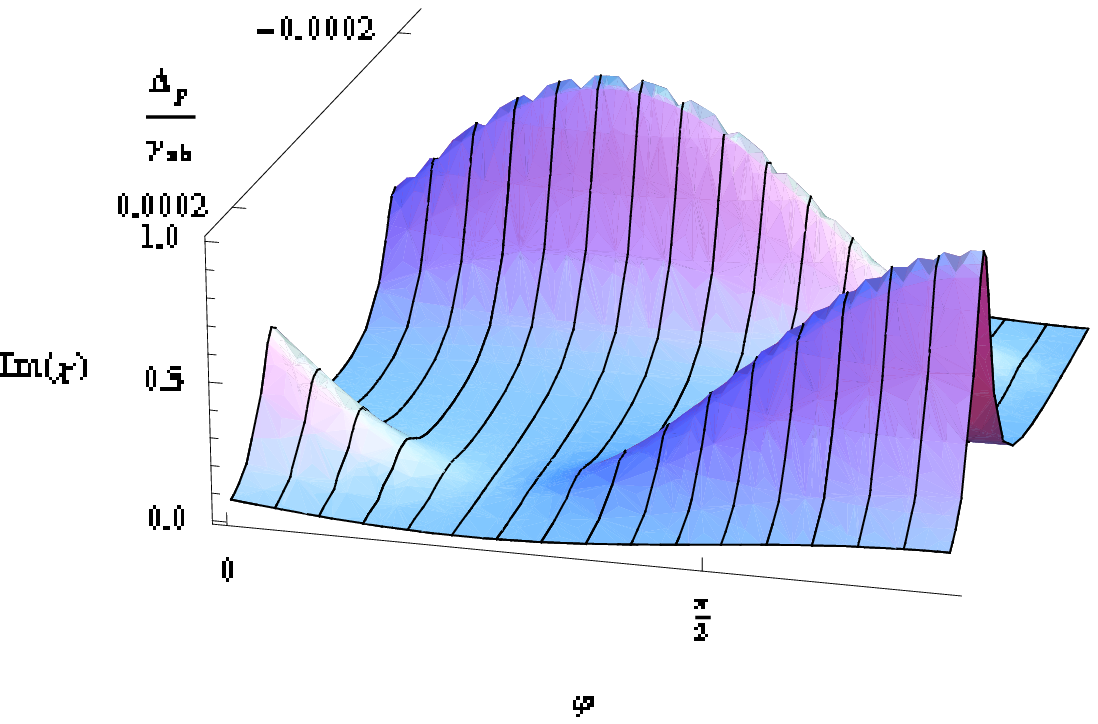}
\caption{\label{varyphi}$\Im[\chi^{(1)}]$ for $g_b=g_c=(2\gamma_{ab})\times10^{-4}$ as a function of $\varphi$.  One can see that the amplitude of each resonance is proportional to the initial population in $|B\rangle$ and $|B'\rangle$.}
\end{figure}

These new resonances can be understood in terms of the interaction of the dressed states of the $|b\rangle$ and $|b'\rangle$ subsystem with the eigenstates of the $|c'\rangle \leftrightarrow |c\rangle \leftrightarrow |a\rangle$ subsystem. To first order in the probe, $|a'\rangle$ can be neglected altogether. First let us consider how the tunnel coupling would effect the probe absorption for the $|b\rangle \rightarrow |a\rangle$ transition without any control laser. The dressed states $|B\rangle$ and $|B'\rangle$ are both populated ground states that couple directly to the excited state by the probe laser. The energies of $|B\rangle$ and $|B'\rangle$ are $\hbar\omega_{B,B'}=\hbar\omega_b\pm g_b^{eff}/2$. Therefore even in the absence of the control laser, the $|b\rangle \rightarrow |a\rangle$ absorption line would be split into two new lines located at $\omega_{a}- \omega_{B}$ and $\omega_a-\omega_{B'}$. This is essentially an Autler-Townes doublet induced by the tunneling. Figure \ref{varyphi} shows the two ultranarrow resonances as a function of $\varphi$, which controls the relative population in the dressed states such that for $\varphi=\pi/4$, $\rho^{(0)}_{BB}=1$ while for $\varphi=3\pi/4$, $\rho^{(0)}_{B'B'}=1$.

The excited state $|a\rangle$ is in fact coupled to $|c\rangle$ via the control laser while $|c\rangle$ is coupled to $|c'\rangle$ via the tunneling. This system is a three-level system that is isomorphic to a $\Lambda$ atom. Again assuming $\delta_{cc'}=0$ and the control laser is on resonance, $\Delta_{\mu}=0$, then we have the following Hamiltonian for the $\{|a\rangle,|c\rangle,|c'\rangle\}$ subsystem,
\begin{equation}
\mathbf{H}=\hbar\omega_a\mathbf{I}+\frac{\hbar}{2}\begin{bmatrix}
0 &\Omega_{ac} &0\\
\Omega_{ac} &0 &-g_c\\
0 &-g_c &0
\end{bmatrix},
\end{equation}
The eigenstates of this Hamiltonian are
\begin{eqnarray}
|a_+\rangle &=&\frac{1}{\sqrt{2}}\left(\sin\theta|a\rangle+|c\rangle+\cos\theta|c'\rangle \right) \\
|a_-\rangle &=&\frac{1}{\sqrt{2}}\left(\sin\theta|a\rangle-|c\rangle+\cos\theta|c'\rangle \right) \\
|a_0\rangle &=& \cos\theta |a\rangle -\sin\theta|c'\rangle
\end{eqnarray}
where $\tan\theta=-\Omega_{ac}/g_c$. The energies of the states $|a_{\pm}\rangle$ are
$E_{\pm}=\hbar\omega_a \pm \hbar\sqrt{\Omega_{ac}^2+g_c^2}/2$
while $|a_0\rangle$ has energy $E_0=\hbar\omega_a$. As one can see, $|a_0\rangle$ is the same type of dark state that appears in STIRAP and coherent population trapping. In this case, this tunneling induced dark state is a superposition of $|a\rangle$ and $|c'\rangle$ {\em but not} $|c\rangle$ . Since it is decoupled from the control laser, there will not be any destructive quantum interference in the probe absorption for transitions to $|a_0\rangle$.
Transitions from the $\{ |B\rangle, |B'\rangle \}$ manifold to $|a_0\rangle$ will then exhibit absorption {\em resonances} at $\omega_a-\omega_{B,B'}$, which correspond to the new ultranarrow resonances. This independence of $|a_0\rangle$ from $|c\rangle$ also explain why the line width, $\Gamma_n$, does not depend on
either $\gamma_{cc'}$ or $\gamma_{bc}$. This is in stark contrast to what would happen if $g_c=0$, which would correspond to a conventional EIT system but with two ground states, $\{ |B\rangle, |B'\rangle \}$. In this case, destructive interference created by the control would lead to nulls in the absorption at $\omega_a-\omega_{B,B'}$.

\begin{figure}
\includegraphics[width=1.0\columnwidth]{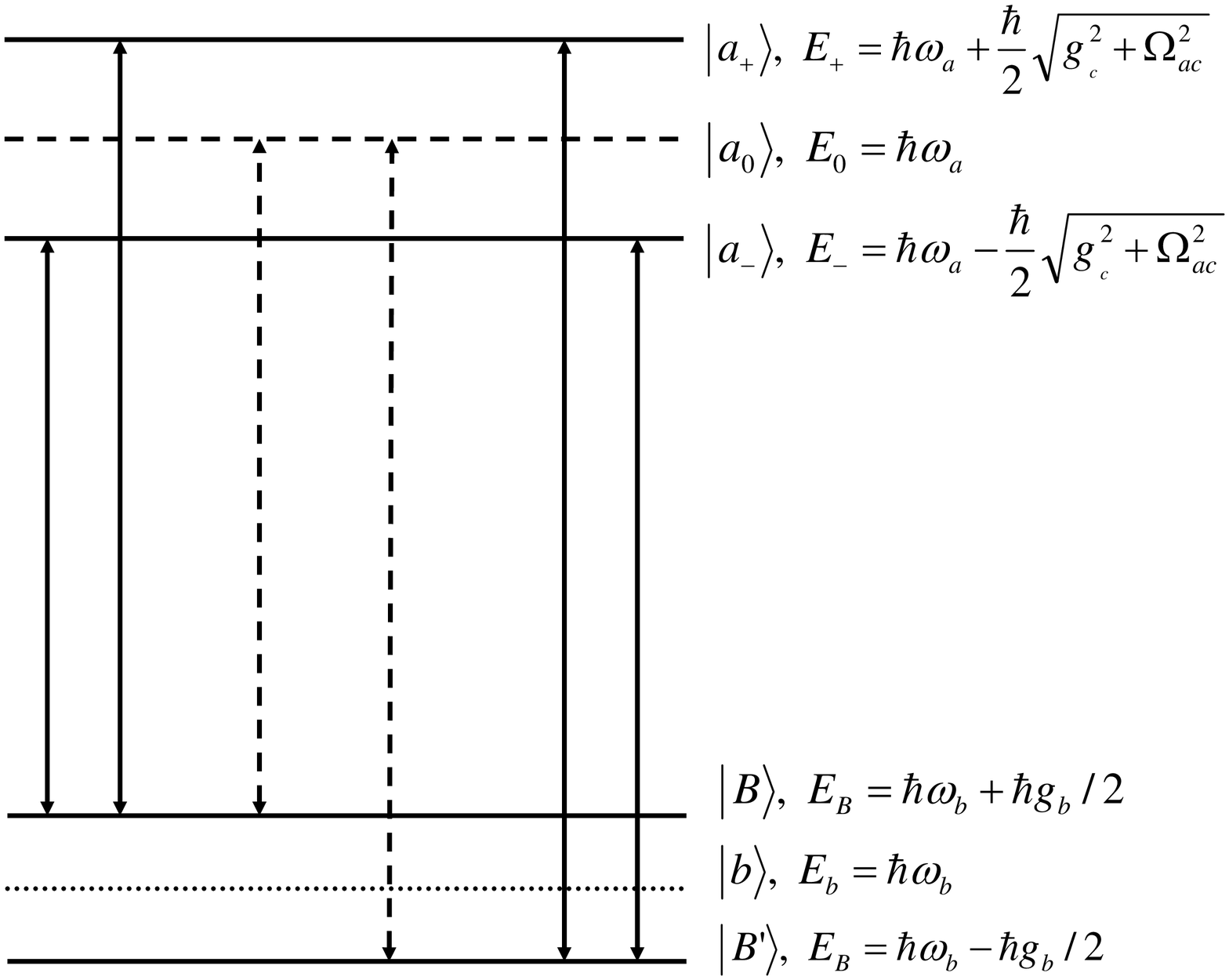}
\caption{\label{dressedstatefig} Energy level diagram that indicates transitions induced by the probe laser between the ground state manifold $\{ |B\rangle, |B'\rangle \}$ and the excited state manifold $\{ |a_+\rangle, |a_-\rangle, |a_0\rangle \}$. Transitions to the dark state $|a_0\rangle$ are indicated by dashed lines. The energy of the bare state $|b\rangle$ is also shown for reference.}
\end{figure}

A general understanding of locations of the absorption resonances can be obtained from Fig. \ref{dressedstatefig} which shows a schematic diagram of the energy levels of the dressed ground state manifold $\{|B\rangle, |B'\rangle \}$, which are coupled to all three states of the excited state manifold $\{ |a_+\rangle, |a_-\rangle, |a_0\rangle \}$ via the probe. All in all there are six transitions that should appear as resonances in the absorption spectrum. The transitions to the two `bright' states $|a_\pm\rangle$ correspond to the main absorption peaks located at $\Delta_p\approx \pm\Omega_{ac}/2$ for $\Omega_{ac}\gg g_c,g_b$. Notice that each of these resonances actually consist of a pair of resonances separated by a distance $g_b$ but because $\gamma_{ab} \gg g_b$ these pairs cannot be individually resolved. Since to first order in the probe field $|a\rangle$ is never populated, the tunneling rate between $|a\rangle$ and $|a'\rangle$, $g_a$ has only a negligible effect on $\chi^{(1)}$.

\begin{figure}
\includegraphics[width=1.0\columnwidth]{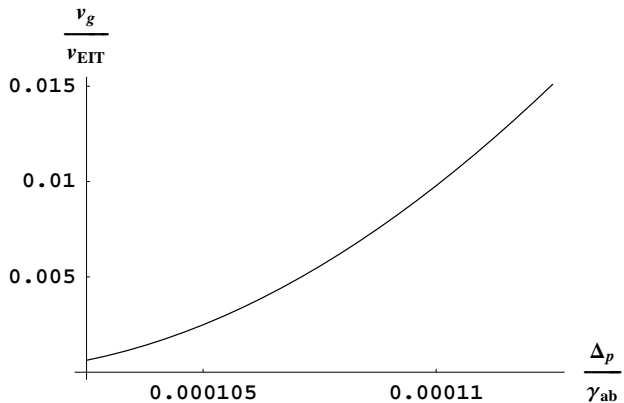}
\caption{\label{fig-groupvelocity}The group velocity in a window of 100Hz just to the side of one of the ultranarrow resonances, in units of the standard minimum EIT group velocity for a given atomic species and control beam strength.  Here $g_b=g_c=(2\gamma_{ab})\times10^{-4}$ and $\varphi=0$.  As can be seen, the group velocity is suppressed by a factor on the order of $10^2$, relative to standard EIT.  The group velocity of standard EIT, $v_{\text{EIT}}$, is calculated in the absence of any tunnel coupling $g_a=g_b=g_c=0$. Absorption is negligible in this frequency window (see Fig. \ref{fig-resonance})}
\end{figure}

As we can see from Eq. \ref{polarization-resonance} there is a transparency window of width $g_b$ in between the two ultranarrow resonances with $\Im[\chi^{(1)}]=0$ at $\Delta_p=0$. To either side of these resonances, the absorption is negligible. In the vicinity of these resonances the dispersion, $\partial \Re[\chi^{(1)}]/\partial \omega_p$, is extremely large and, to either side of the peak, there is a region of width $\approx g_c/2$ in which the dispersion is 10 times greater than in standard EIT while within a region of width $\approx g_c/8$, the dispersion is amplified by a factor of 100.  (Note that `Standard EIT' refers to the case where $g_a=g_b=g_c=0$ but with all other parameters, including the control laser, being the same.)
The absorption, meanwhile, drops to below $1\%$ within an order of magnitude of the feature's width---$2(g_c/\Omega_{ac})^2\gamma_{ab}$---from the center of the resonance, and is negligible within the regions of interest.  Using our approximation of $g_c\approx 1$KHz, we find there is a region, to either side of the peaks, of width $\mathcal{O}( 1$KHz$)$ in which the absorption is negligible ($\lesssim .001\%$) and the dispersion is $\approx$ 10 times greater than in standard EIT with a control laser of the same strength;  similarly there is a region of width $\mathcal{O}(100$Hz$)$ in which the dispersion is $\approx$ 100 times greater than standard EIT, again with negligible absorption (likewise $\lesssim .001\%)$. This implies that the group velocity \cite{fleischhauer-review,hau-first-observation},
\[
v_g(\omega_p)=\frac{c}{n+(\omega_p/2n)(\partial\Re[\chi^{(1)}]/\partial \omega_p)}
\]
could be made significantly smaller than in previous experiments (see Fig. \ref{fig-groupvelocity}). Note that the slope of the dispersion is rapidly changing within each of these regions, and so any pulse transmitted through the well would undergo considerable reshaping.

\section{Discussion and Conclusions}\label{experimental_prospects}

The regions in which dispersion is especially high in this system are on the order of $g_c$ or smaller, begging the question of their experimental accessibility using readily available equipment.  We propose two solutions.  The first follows a recent paper by Pfau's group, at Stuttgart \cite{pfau}.  In their experiment, only a $\sigma^+$ polarized probe couples the initial state of the atoms to the same excited state as the control laser. On the other hand, a $\sigma^-$ polarized probe couples off-resonantly to other excited states that are unaffected by the control laser. By mixing a small amount of $\sigma^-$ polarized light into their otherwise $\sigma^+$ polarized probe beam, they simultaneously measured the absorption and dispersion of the EIT system by examining the interference pattern between the $\sigma^+$ and $\sigma^-$ polarized components of the probe. They report observing features as narrow as 4KHz in the $\Re[\chi^{(1)}]$.  Performing a similar experiment with a system prepared as described in this paper would test our predictions for the modified absorption and dispersion arising from the presence of the second well.

To directly observe the low group velocity that follows from our predictions would require lasers with line widths small relative to the frequency window in which the dispersion is large.  Though such lasers are not readily available commercially, several groups have reported performing spectroscopic experiments within the range of interest.  As early as 1999, for instance, Young et al. at NIST were able to achieve sub-hertz width lasers, albeit with nontrivial active stabilization \cite{nist-subhertz}.  More recently, using a a 657-nm diode laser with a femtosecond comb, a collaboration at NIST and LANL performed KHz-resolution spectroscopy on cold neutral calcium \cite{nist-kilohertz}.  And so, the next generation of lasers could take full advantage of the high-dispersion regime exhibited by our double well system.

Finally we would like to comment on the approximations made in our model of a double well BEC that could affect the experimental feasibility of our proposal.
The major assumption made is that the spatial profile of the condensate is fixed. This means that the ground state wave functions $u^{(L/R)}_\ell ({\bf r})$ are unchanging and also that there are no excitations of the condensate. Similar `two-mode' models for a double well BEC have been successfully used to obtain quantitative agreement with experiment \cite{bergmann, gati-noise-thermometry,gati-noise-thermometry-2}.

In the limit of small nonlinear interactions between atoms, $u^{(L/R)}_\ell ({\bf r})$ are the single particle wave functions, Gaussians in the case of our harmonic potential. The shape of these wave functions are not affected by the population in the wells. However, in the limit of large nonlinear interactions, $u^{(L/R)}_\ell ({\bf r})$ are best approximated by Thomas-Fermi wave functions and therefore will only remain the same if the number of atoms in each well does not change with time. This is implicit in our choice of initial conditions and a weak probe field that does not significantly excite $|a\rangle$. Furthermore, the dynamics of $|a\rangle$ are dictated by the laser coupling, which is faster than the time needed for atoms in $|a\rangle$ to equilibrate in the potential. Atoms excited from $|b\rangle$ to $|a\rangle$ will not have time to equilibrate in the new potential and therefore $u^{(L/R)}_a ({\bf r})=u^{(L/R)}_b ({\bf r})$ implying that the prefactor in Eq. \ref{genpol} is maximal: $[u_a(\mathbf{r})]^*u_b(\mathbf{r})\rightarrow |u_b(\mathbf{r})|^2$.

The second issue is that of excitations of the condensate. Transverse excitations can easily be avoided by using tight confinement in the x and y directions ($\omega_x,\omega_y>10^3 s^{-1}$). However, since we want to maximize the optical thickness of the wells along the z direction, we must have very weak confinement along this axis, implying that excitations along this direction could readily be excited. These long wavelength excitations should have a negligible impact for two reasons. First, the measured signal is proportional to the integral of the condensate density in the z-direction. Even if there are local spatial variations of the density in the z-direction, those variations would be averaged out in the integration. Secondly, the trapping frequency along the z-axis will be at least one to two orders of magnitude smaller than in the x and y directions ($\omega_z\approx10s^{-1}-100s^{-1}$). The momentum distribution of such elongated quasi-1D condensates have been measured experimentally using Bragg spectroscopy \cite{richard-momentum} and showed to have a momentum distribution in the axial direction ranging from $50Hz$ to $500Hz$ for temperatures ranging from $T=0.25T_C$ to $T=0.9T_C$ and trapping potentials $\omega_x=\omega_y=4700s^{-1}$ and $\omega_z=30s^{-1}$ ($T_C$ is the critical temperature for Bose-Einstein condensation). Since experiments have already been performed with interwell tunnel couplings as high as $7900Hz$ \cite{gati-noise-thermometry}, this would imply that at low temperatures the separation between the ultranarrow resonances would be more than 100 times greater than the inhomogeneous broadening in the z-direction. Additionally we can point to the work in Ref. \cite{gati-noise-thermometry} that measured the affect of thermal excitations in a double well BEC Josephson junction. This work showed that the loss of inter-well
phase coherence due to thermal excitations was negligible when the tunnel coupling energy was much larger than the thermal energy, $k_BT$.
Therefore we conclude that for $T\ll T_C$, inhomogeneous broadening due excitations will be sufficiently small as to not affect our results.

In conclusion, we have studied the effect of interwell tunneling on the EIT dispersion and absorption of a probe laser interacting with a double well Bose-Einstein condensate. As a future direction we plan to extend these results to an optical lattice where the ultranarrow resonances should be determined by the band structure of the lattice. Additionally, the effect of tunneling on the $\chi^{(3)}$ nonlinear susceptibility will be explored.

\section{Appendix A}
Here we first present the Gross-Pitaevskii equations in terms of the dressed states of the subspaces $\{\ket{b},\ket{b'}\}$ and $\{\ket{c},\ket{c'}\}$
\begin{subequations}
\begin{eqnarray}
\I\dtime{\tilde{\psi}_a}&=&\left(\omega_a-\omega_p-\omega_{\mu}+\tilde{\Psi}^{\dagger}\mathbf{U}_a\tilde{\Psi}\right)\tilde{\psi}_a-\frac{g_a}{2}\tilde{\psi}_{a'}\nonumber\\
&-&\frac{\Omega_{ab}}{2}e^{-\I\phi_{ab}}\left(\cos{\theta_b}\tilde{\psi}_B-\sin{\theta_b}\tilde{\psi}_{B'}\right)\nonumber\\
&-& \frac{\Omega_{ac}}{2}e^{-\I\phi_{ac}}\left(\cos{\theta_c}\tilde{\psi}_C-\sin{\theta_c}\tilde{\psi}_{C'}\right)\\
\I\dtime{\tilde{\psi}_B}&=&\left(\omega_{b}-\omega_{\mu}-\frac{g_b^{\text{eff}}}{2}+\frac{1}{2}(\tilde{\Psi}^{\dagger}\mathbf{U}_b\tilde{\Psi} +\tilde{\Psi'}^{\dagger}\mathbf{U}_b\tilde{\Psi'})\right)\tilde{\psi}_B\nonumber\\
&-&\frac{\Omega_{ab}}{2}e^{\I\phi_{ab}}\cos{\theta_b}\tilde{\psi}_a\\
\I\dtime{\tilde{\psi}_C}&=&\left(\omega_{c}-\omega_{p}-\frac{g_c^{\text{eff}}}{2}+\frac{1}{2}(\tilde{\Psi}^{\dagger}\mathbf{U}_c\tilde{\Psi} +\tilde{\Psi'}^{\dagger}\mathbf{U}_c\tilde{\Psi'})\right)\tilde{\psi}_C\nonumber\\
&-&\frac{\Omega_{ac}}{2}e^{\I\phi_{ac}}\cos{\theta_c}\tilde{\psi}_a
\end{eqnarray}
\begin{eqnarray}
\I\dtime{\tilde{\psi}_{a'}}&=&\left(\omega_a-\omega_{\mu}-\omega_p+\tilde{\Psi'}^{\dagger}\mathbf{U}_a\tilde{\Psi'}\right)\tilde{\psi}_{a'}-\frac{g_a}{2}\tilde{\psi}_a\\
\I\dtime{\tilde{\psi}_{B'}}&=&\left(\omega_{b}-\omega_{\mu}+\frac{g_b^{\text{eff}}}{2}+\frac{1}{2}(\tilde{\Psi}^{\dagger}\mathbf{U}_b\tilde{\Psi} +\tilde{\Psi'}^{\dagger}\mathbf{U}_b\tilde{\Psi'})\right)\tilde{\psi}_{B'}\nonumber\\
&+&\frac{\Omega_{ab}}{2}e^{\I\phi_{ab}}\sin{\theta_b}\tilde{\psi}_a\\
\I\dtime{\tilde{\psi}_{C'}}&=&\left(\omega_{c}-\omega_{p}+\frac{g_c^{\text{eff}}}{2}+\frac{1}{2}(\tilde{\Psi}^{\dagger}\mathbf{U}_c\tilde{\Psi} +\tilde{\Psi'}^{\dagger}\mathbf{U}_c\tilde{\Psi'})\right)\tilde{\psi}_{C'}\nonumber\\
&+&\frac{\Omega_{ac}}{2}e^{\I\phi_{ac}}\sin{\theta_c}\tilde{\psi}_a.
\end{eqnarray}
\end{subequations}
The rotation angle $\theta_b$ and $g_b^{eff}$ are defined in the text. The transformation to the dressed state basis $\{|C\rangle,|C'\rangle\}$ from $\{\ket{c},\ket{c'}\}$ is defined in the same manner as Eq. (10) with the rotation angles and dressed state energies explicitly given by
\begin{subequations}
\begin{eqnarray}
 &\cos\theta_c = \left(\frac{1-\delta_{cc'}/g^{\text{eff}}_{c}}{2}\right)^{1/2}\\
&\sin\theta_c = \left(\frac{1+\delta_{cc'}/g^{\text{eff}}_{c}}{2}\right)^{1/2}\\
 &g_c^{\text{eff}}=\sqrt{\delta_{cc'}^2+g_c^2}.
\end{eqnarray}
\end{subequations}
and $\delta_{cc'}=\left(\tilde{\Psi}^{\dagger}\mathbf{U}_c\tilde{\Psi}-\tilde{\Psi'}^{\dagger}\mathbf{U}_c\tilde{\Psi'}\right)/2$.

\section{Appendix B}
The general solution for the coherence $\tilde{\rho}_{ab}$ is given by the steady state solution of Eqs. (\ref{system1}) and (\ref{system2}) combined with Eq. (\ref{rhoab}). The general form of the steady state $\tilde{\rho}_{ab}$ is then:
\begin{widetext}
\begin{align}\label{fullnumerator}
\tilde{\rho}_{ab}&=e^{-i \phi_{ab}} \Omega_{ab}\left(\frac{\cos^2(\theta_b-\varphi)\cos^2\theta_b}{\zeta_-}\left(\left(2\Delta_{\mu}-2\Delta_{p}-g_b^{\text{eff}}\right)^2-(g_c^{\text{eff}})^2\right) \left(2\Delta_p+g_b^{\text{eff}}-2i\gamma_{ab}\right)\right.\notag\\
&+\left.\frac{\sin^2(\theta_b-\varphi)\sin^2\theta_b}{\zeta_+}\left(\left(2\Delta_{\mu}-2\Delta_{p}+g_b^{\text{eff}}\right)^2-(g_c^{\text{eff}})^2\right) \left(2\Delta_p-g_b^{\text{eff}}-2i\gamma_{ab}\right)\right)
\end{align}
while the population in $|a\rangle$ to second order in the probe field is:
\begin{align}
    \tilde{\rho}_{aa}&=\frac{i\Omega_{ab}^2}{\gamma_a}\left( \left(\frac{\cos^2(\theta_b-\varphi)\cos^2\theta_b}{\zeta_-}\left(\left(2\Delta_{\mu}-2\Delta_{p}-g_b^{\text{eff}}\right)^2-(g_c^{\text{eff}})^2\right) \left(2\Delta_p+g_b^{\text{eff}}-2i\gamma_{ab}\right)\right.\right. \nonumber\\
&+\left.\frac{\sin^2(\theta_b-\varphi)\sin^2\theta_b}{\zeta_+}\left(\left(2\Delta_{\mu}-2\Delta_{p}+g_b^{\text{eff}}\right)^2-(g_c^{\text{eff}})^2\right) \left(2\Delta_p-g_b^{\text{eff}}-2i\gamma_{ab}\right)\right) \nonumber \\
&- \left(\frac{\cos^2(\theta_b-\varphi)\cos^2\theta_b}{\zeta_-^*}\left(\left(2\Delta_{\mu}-2\Delta_{p}-g_b^{\text{eff}}\right)^2-(g_c^{\text{eff}})^2\right) \left(2\Delta_p+g_b^{\text{eff}}+2i\gamma_{ab}\right)\right.\nonumber\\
&+\left.\left.\frac{\sin^2(\theta_b-\varphi)\sin^2\theta_b}{\zeta_+^*}\left(\left(2\Delta_{\mu}-2\Delta_{p}+g_b^{\text{eff}}\right)^2-(g_c^{\text{eff}})^2\right) \left(2\Delta_p-g_b^{\text{eff}}+2i\gamma_{ab}\right)\right)\right)
\end{align}
Here we have defined
\begin{align}\label{fulldenominator}
\zeta_{\pm}&=\left(\left(2\Delta_{\mu}-2\Delta_p\pm g_b^{\text{eff}}\right)^2-(g_c^{\text{eff}})^2\right)\left(\left(2\Delta_p\mp g_b^{\text{eff}}-2i\gamma_{ab}\right)^2-g_a^2\right)\notag\\
&+\left(2\Delta_p\mp g_b^{\text{eff}}-2i\gamma_{ab}\right)\left(2\Delta_{\mu}-2\Delta_p\pm g_b-g_c\cos(2\theta_c)\right)\Omega_{ac}^2
\end{align}
We have simplified equations (\ref{fullnumerator}) through (\ref{fulldenominator}) by making the substitutions $\Delta_p-\frac{U_{bb}}{2}+U_{ab}\rightarrow\Delta_p$ and $\Delta_{\mu}-\frac{U_{cb}}{2}+U_{ab}\rightarrow\Delta_{\mu}$.
\end{widetext}

\bibliography{semiclassical}

\end{document}